\documentclass{ws-rv975x65}
\usepackage{subfigure}     % required only when side-by-side / subfigures are used
\usepackage{ws-rv-thm}     % comment this line when `amsthm / theorem / ntheorem` package is used
\usepackage{ws-rv-van}     % numbered citation/references (default)
\makeindex
\begin{document}

\chapter[X-Ray Polarimetry]{X-Ray Polarimetry}\label{ch_xraypolarimetry}

\author[P. Kaaret]{Philip Kaaret}

\address{Department of Physics and Astronomy, University of Iowa \\
Van Allen Hall, Iowa City, IA 52242, USA \\
philip-kaaret@uiowa.edu}

\begin{abstract}

We review the basic principles of X-ray polarimetry and current detector technologies based on the photoelectric effect, Bragg reflection, and Compton scattering.  Recent technological advances in high-spatial-resolution gas-filled X-ray detectors have enabled efficient polarimeters exploiting the photoelectric effect that hold great scientific promise for X-ray polarimetry in the 2--10~keV band.  Advances in the fabrication of multilayer optics have made feasible the construction of broad-band soft X-ray polarimeters based on Bragg reflection.  Developments in scintillator and solid-state hard X-ray detectors facilitate construction of both modular, large area Compton scattering polarimeters and compact devices suitable for use with focusing X-ray telescopes.

\end{abstract}

%\markright{Customized Running Head for Odd Page} % default is Chapter Title.
\body

\section{Polarization}

The polarization of photons reflects their fundamental nature as electromagnetic waves.  A photon is a discrete packet of electric and magnetic fields oriented transverse to the direction of motion.  The fields evolve in time and position according to Maxwell's equations.  The polarization describes the configuration of the fields.  Since the electric and magnetic fields are interrelated by Maxwell's equations, the configuration of both fields is set by specification of the electric field alone.  

An electromagnetic plane wave propagating along the $z$-axis with angular frequency $\omega$ can be described as a sinusoidally varying electric field of the form

\begin{equation}
\vec{E} = \hat{x} E_{X} + \hat{y} E_{Y} = 
          \hat{x} E_{0X} \cos(kz - \omega t) +
          \hat{y} E_{0Y} \cos(kz - \omega t + \xi).
\end{equation}

\noindent where $\xi$ and the ratio of $E_{0X}$ versus $E_{0Y}$ set the polarization and the wavenumber $k = \omega / c$.  Polarization is symmetric under a 180$^{\circ}$ rotation, since such a rotation can be produced by translation in time or space.   The wave is linearly polarized if $E_{X}$ and $E_{Y}$ are always proportional.  This occurs when $\xi = n \pi$, where $n$ is an integer, so that $E_{X}$ and $E_{Y}$ are exactly in phase or antiphase.  The polarization angle is set by the ratio of $E_{0Y}$ and $E_{0X}$.  If $\xi \ne n \pi$, the electric field rotates as a function of time or position, which is elliptical polarization.  The polarization is described as right or left handed according to whether $\vec{E}$ rotates clockwise or counterclockwise.  Circular polarization is the special case of elliptical polarization with $E_{0X} = E_{0Y}$.

Each individual photon is necessarily polarized.  However, different photons from a particular source may have different polarizations.  If the distribution of polarization angles is uniform, then the source has zero net polarization.  If not, then the source has a net polarization.    Generation of non-zero net polarization requires a net deviation from spherical symmetry in either the physical geometry or the magnetic field configuration of the astrophysical system.

The Stokes parameters provide a means to fully characterize the polarization of a source using four intensities: 
\begin{eqnarray}
I = \langle E_{0X}^2 \rangle + \langle E_{0Y}^2 \rangle \\
Q = \langle E_{0X}^2 \rangle - \langle E_{0Y}^2 \rangle \\
U = \langle 2 E_{0X} E_{0Y} \cos \xi \rangle \\
V = \langle 2 E_{0X} E_{0Y} \sin \xi \rangle
\end{eqnarray}

\noindent The averages are taken over the photons detected from the source.  The fractional degree of polarization, also called the polarization fraction or the magnitude of polarization, is $P = \sqrt{Q^2 + U^2 + V^2}/I$.  The polarization position angle is $\tan (2 \phi_0) = U/Q$.  The Stokes parameter $V$ describes elliptical polarization.  Since most X-ray polarimeters are sensitive only to linear polarization, we will not consider elliptical polarization further.

\section{Polarization Measurement}

Available X-ray instrumentation is able to measure the intensity of X-rays (the number of photons per unit time), the energies of X-rays (via conversion of that energy to charge or heat), and the positions of X-rays or, more precisely, the positions at which an X-ray deposits charge via interactions.  Since X-ray polarization cannot be measured directly, the X-rays must first undergo some interaction that converts the polarization information to a directly measurable quantity, typically intensity or position.\cite{Novick75}

\begin{figure}[ht]
% one figure environment to get the two figures on same page

\centerline{\includegraphics[width=4in]{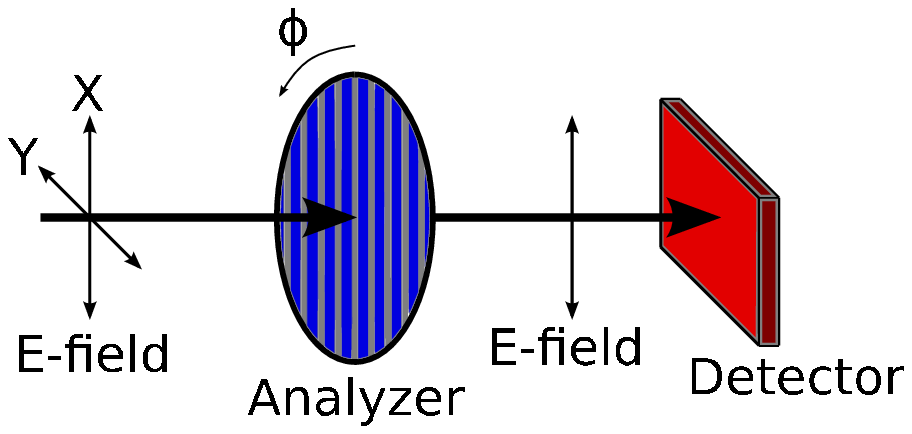}}
\caption{Polarization analyzer.  A linear polarization analyzer is rotated and the associated detector records the intensity of photons (counts) at each angle as a modulation curve as shown in Figure~\ref{modcurve}.}
\label{analyzer} \bigskip

\centerline{\includegraphics[width=4in]{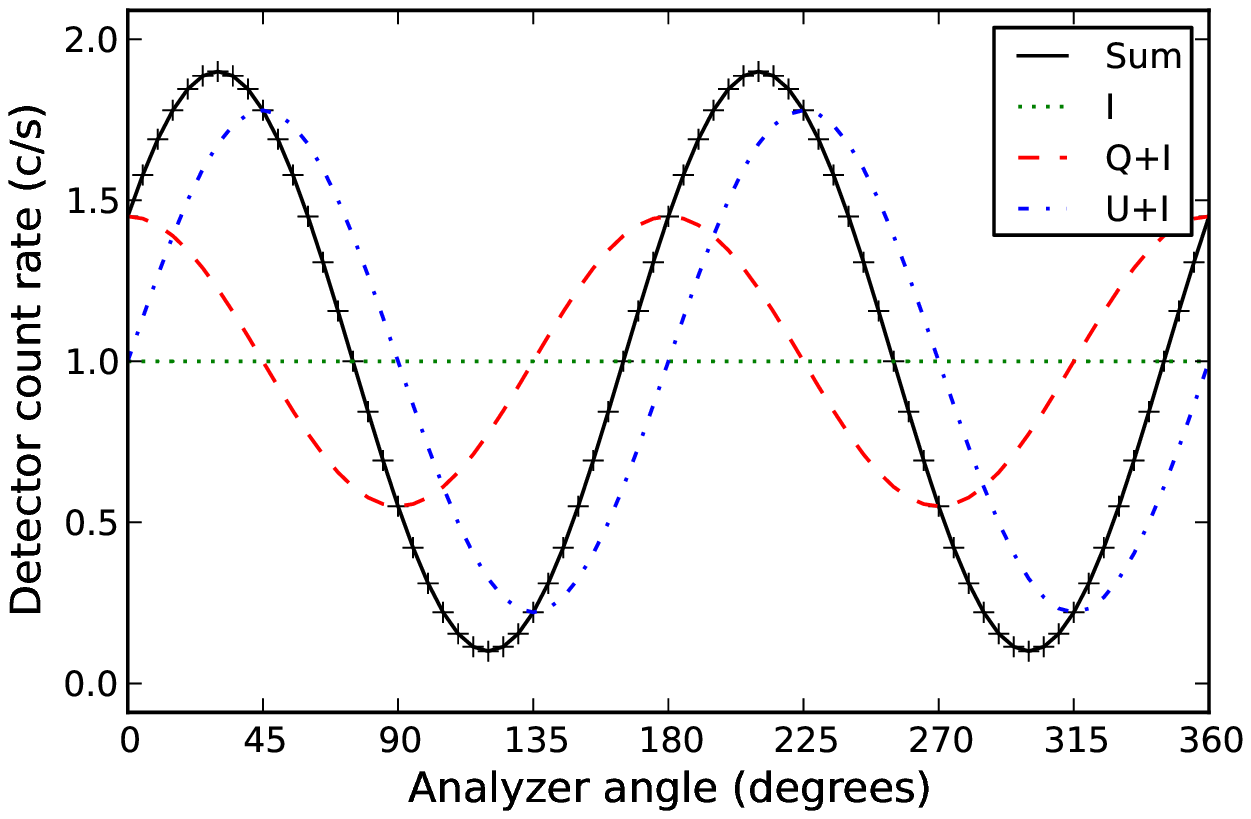}}
\caption{Modulation curve: detector count rate versus rotation angular of a linear polarizing filter.  The crosses indicate data points.  The dashed curves are sums of the Stokes decomposition as indicated.  The solid curve is the sum of the three Stokes components.  The curve has $a = 0.9$ and $\phi_0 = 30^{\circ}$.}
\label{modcurve}

\end{figure}

In Figure~\ref{analyzer}, we consider one rotating linear polarization analyzer.  As the analyzer is rotated, the associated detector records the intensity of photons (counts) at each analyzer angle.  The resulting histogram of counts versus rotation angle, or modulation curve, is shown in Figure~\ref{modcurve}.  In general, the modulation curve will have the form
\begin{equation}
S(\phi) = A + B \cos^2(\phi - \phi_0)
\end{equation}

\noindent The polarization position angle $\phi_0$ is the angle at which the maximum intensity is recorded, $A$ describes the unpolarized component of the intensity, and $B$ describes the polarized intensity.  The modulation amplitude is $a = (S_{max} - S_{min})/(S_{max} + S_{min}) = B/(2A + B)$.  Given a modulation curve, $a$ and $\phi_0$ can be obtained by non-linear regression.

The modulation curve can also be written in terms of the Stokes parameters as
\begin{equation}
S(\phi) = I + Q \cos(2\phi) + U \sin(2\phi)
\end{equation}

\noindent The Stokes decomposition is equivalent to a Fourier series with one period.  The Stokes parameters can be obtained directly from the modulation curve: $I = \langle S(\phi) \rangle$, $Q = S(0^{\circ}) - I$, and $U = S(45^{\circ}) - I$.  This can be visually verified in Figure~\ref{modcurve} as $Q + I = S(0^{\circ})$, where the $U$ sinusoid is zero, and $U + I = S(45^{\circ})$, where the $Q$ sinusoid is zero.  This determination of the Stokes parameters is equivalent to measuring the source intensity through three filters\footnote{Polarization measurements are frequently done with such sets of filters in the optical/IR.}: unpolarized, polarized at $0^{\circ}$, polarized at $45^{\circ}$.  A key feature of the Stokes decomposition is that the modulation curve is linear in the Stokes parameters, thus they can be obtained via {\it linear} regression \cite{Strohmayer13}.  The magnitude, $P$, and angle, $\phi_0$ of polarization can be recovered from the Stokes parameters using the equations in the previous section or fit for directly from the modulation curve.  The Stokes parameters are particularly useful because they are additive if fitting multiple modulation curves.  This is not true of the magnitude and angle of polarization.

The discussion above assumes the use of an ideal polarization analyzer that passes no radiation if oriented perpendicular to a 100\% polarized beam.  In this case, the modulation amplitude is equal to the polarization fraction.  The real world is less than perfect.  Most actual polarization analyzers pass some fraction of the radiation even when oriented perpendicular to a 100\% polarized beam.  The `modulation factor', $\mu$, is defined as the modulation amplitude measured by a polarimeter for a 100\% polarized beam.  The modulation factor is a property of the polarimeter and may also depend on the energy or spatial distribution of the input photons.  Background events (events not produced by photons from the source) also dilute the modulation curve.  For a polarimeter with a measured $\mu$ and background count rate $b$ independent of rotation angle, the polarization fraction of a source that produces a modulation amplitude $a$ and an average count rate $r$ is 
\begin{equation}
P = \frac{a}{\mu} \frac{r+b}{r}.
\end{equation}

In designing an X-ray polarimeter, it is essential that the system (analyzer/detector and telescope) can reach sufficient statistical accuracy for the measurements required.  The traditional figure of merit is the `Minimum Detectable Polarization' (MDP).\cite{Novick77}.  The polarization fraction, $P$, is a non-negative quantity.  Thus, due to statistical fluctuations, any particular measurement of $P$ will produce a value greater than 0.  The MDP is the largest fluctuation expected to occur with a probability of 1\%.  Equivalently, the MDP is the smallest polarization that can be detected at a 99\% confidence level.  The MDP for an observation of duration $T$ is
\begin{equation}
{\rm MDP} = \frac{4.29}{\mu r} \sqrt{\frac{r+b}{T}} 
          = \frac{4.29}{\mu} \frac{1}{\sqrt{N}} \sqrt{1+\frac{b}{r}},
\end{equation}

\noindent where $N = rT$ is the total number of source counts.  Reaching an MDP of 1\% with an ideal polarimeter, $\mu = 1$ and $b = 0$, requires $\sim$200,000 counts.

\begin{figure}[ht]
\centerline{\includegraphics[width=3in]{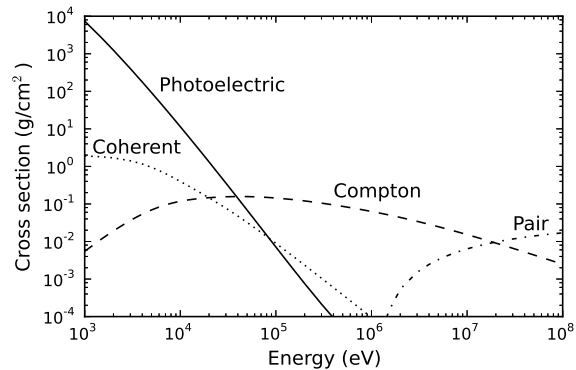}}
\caption{Photon interactions: mass attentuation coefficients for photoelectric (solid line), coherent scattering (dotted line), Compton scattering (dashed line), and pair production (dash-dot line) interactions of photons in neon versus energy \cite{Berger98}. }
\label{cross}
\end{figure}

Usually, a scientifically useful polarization measurement entails determination of both $P$ and $\phi_0$.  This is a joint measurement of two parameters and requires additional statistics beyond those suggested by the MDP \cite{Weisskopf10}.  For small polarization amplitudes and $b/r \ll 1$, an increase in counts by a factor $\sim 2.2$ is needed to maintain a 99\% joint confidence interval for two parameters \cite{Weisskopf10,Strohmayer13}.  The factor decreases as the polarization amplitude increases.

The X-ray polarization levels predicted for astronomical objects are often quite low, near 1\%, thus instrumental or systematic errors are a serious concern.  Accurate calibration, including with unpolarized beams, is essential for successful polarization measurements.\cite{Weisskopf10rome}  Also, rotation of the instrument is a powerful tool to understand and remove the effects of systematic errors.  The fact that polarization is symmetric under a 180$^{\circ}$ rotation can also be used to check for systematic errors, even for polarimeters that require rotation to perform the measurement.  Since most astronomical X-ray sources are time varying, the rotation period should either be shorter than the typical time scale of variability, or many rotations should be executed during each individual observation.

\section{Physical Processes for Polarization Measurement}

The mass attentuation coefficients for interaction of photons with neon is shown as a function of energy in Figure~\ref{cross}. Photoelectric interactions dominate at low energies, Compton scattering dominates at intermediate energies, and pair production dominates at the highest energies.  The mass attentuation coefficients are similar for other elements, but the transitions shift to higher energies for higher atomic number.  The mass attentuation coefficient determines which interaction is most effective for polarization analysis in each band: photoelectric below a few tens of keV and Compton in the hard X-ray/soft gamma-ray band.  Bragg reflection (coherent scattering from a crystal or multilayer) has been used for X-ray polarimetry in the `standard' X-ray band from 2--10~keV and demonstrates promise in the soft X-ray band.  
%The polarization dependence of specular reflection can also be exploited and will be strongest at low energies where X-rays can be reflected through sizable graze angles.

The design of an X-ray polarimeter depends strongly on the physical interaction used to obtain polarization sensitivity.  In the following sections, we review current work on X-ray polarimeters exploiting different physical processes used for polarization analysis.

\begin{figure}[tb]
\centerline{\includegraphics[width=3in]{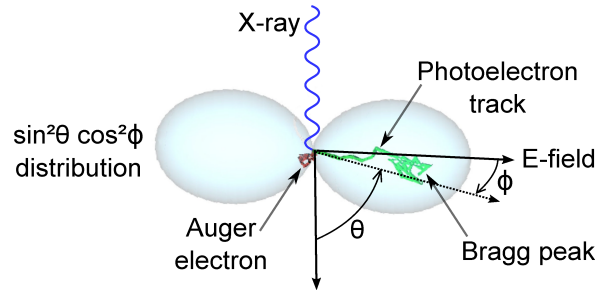}}
\caption{Angular distribution of the photoelectron emitted by interaction of a linearly polarized photon with an atom.  The photoelectron is emitted preferentially along the photon electric field, but not necessarily exactly parallel to the electric field.  The direction of emission is described by two angles: $\phi$ is the azimuthal angle relative to the photon electric field vector, $\theta$ is the emission angle relative to the photon momentum vector.}
\label{photoeffect}
\end{figure}

\section{Photoelectric X-Ray Polarimeters}

\subsection{Photoelectric interaction}

In a photoelectric interaction between an X-ray and an atom, an electron (the `photoelectron') is ejected from an inner shell of an atom with a kinetic energy equal to the difference between the photon energy and the binding energy.  The photoelectron direction is determined by the electric field of the photon.  For a linearly polarized photon, the photoelectron angular distribution is given by
\begin{equation}
\frac{d\sigma}{d\Omega} = \frac{\sin^2(\theta) \cos^2(\phi)}
                               {(1-\beta \cos(\theta))^4}
\end{equation}

\noindent where $\phi$ is the photoelectron azimuthal angle relative to the photon electric field vector, $\theta$ is the photoelectron emission angle relative to the photon momentum vector, and $\beta$ is the photoelectron speed as a fraction of the speed of light (see Fig.~\ref{photoeffect}). For low energy photons (up to tens of keV), leading to low energy electrons and $\beta \ll 1$, the photoelectron is emitted preferentially in the plane perpendicular to the photon momentum vector, $\theta = 90^{\circ}$.  For more energetic photons and photoelectrons, the distribution shifts toward the forward direction.

The photoelectron is preferentially emitted parallel to the photon electric field, i.e.\ the distribution peaks at $\phi = 0^{\circ}$.  Thus, it is possible to determine the linear polarization of the incident photon by measuring the initial direction of the photoelectron.  The photoelectric effect is an ideal polarization analyzer -- the probability of ejecting a photoelectron perpendicular to the electric field vector is zero.

\begin{figure}[tb]
\centerline{\includegraphics[width=4in]{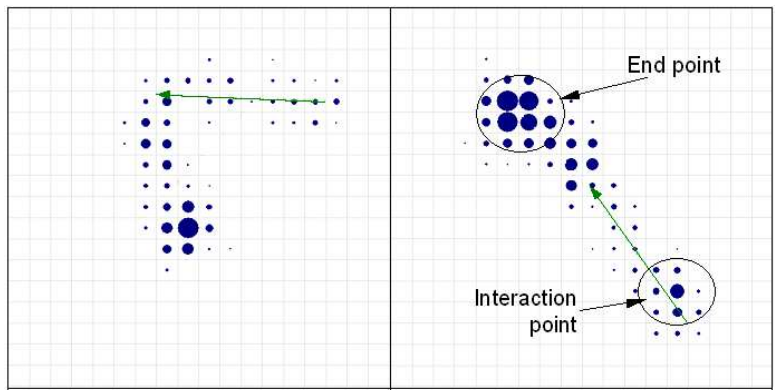}}
\caption{Photoelectron track.  The image on the right shows a relatively straight track with the interaction point, end point, and initial photoelectron direction marked.  The image on the left shows a track where the photoelectron has suffered substantial scattering, from Ref.~\refcite{Black07}.}
\label{petrack}
\end{figure}

\subsection{Photoelectron track}

Once the photoelectron is emitted, it interacts with the surrounding matter.  The photoelectron ionizes atoms, producing electron-ion pairs and changing its own direction and losing energy.  It also scatters off atomic nuclei, changing its direction but with no significant energy loss.  The photoelectron leaves a trail of electron-ion pairs marking its path from initial ejection to final stopping point.  This trail is referred to as the photoelectron `track'.\cite{Costa01}

A photoelectron track is shown in Figure~\ref{petrack}.  Since the photoelectron is emitted preferentially in the plane perpendicular to the photon momentum vector, it is usually sufficient to reconstruct the photoelectron track only in that plane.  To extract the initial direction of the photoelectron one must: 1) determine which is the starting end of the track, 2) measure the angle of the track near its start.  The energy loss rate (per distance traveled) of the photoelectron is inversely proportional to its instantaneous energy.\cite{Soffitta01}  Thus, the energy loss is lowest near the initial part of the track and highest at the end.  The concentrated energy loss near the end of the track is the `Bragg peak'.  This asymmetry in energy loss provides a means to identify the start versus end of the track.  Once the start of the track is identified, one must then fit some portion of the track profile to reconstruct the initial photoelectron direction.  Because the photoelectron scatters as it moves through the gas, the track is not straight.  Minimizing the track length used for the initial direction fitting minimizes the effect of scattering.  However, a sufficient track length must be used to obtain an accurate measurement of the initial direction, since the track has a non-zero width due to electron diffusion and detector resolution and also since the statistical accuracy improves with the number of secondary electrons used.  Thus, the track reconstruction algorithm must balance these factors.\cite{Bellazzini03}

Another complication in track fitting arises from `Auger electrons'.  The ejection of a photoelectron leaves the atom with an unfilled orbital, often in a core shell.  The orbital is refilled by an outer shell electron accompanied with emission of a photon or an electron, necessary for energy conservation.  Emission of a fluorescence photon usually does not affect the photoelectron track, since the photon absorption length is long compared to the track length.  However, emission of an electron complicates the photoelectron track, leading to a reduction in the modulation factor.  Electrons are emitted via the Auger process in which one outer shell electron fills the core orbital, while a second outer shell electron is emitted, leaving the atom doubly ionized.  The Auger electron energy is equal to the difference between the binding energy of the core orbital and the sum of energies of the two outer orbitals.  The probability for Auger emission is high for elements with low atomic number.  However, use of low atomic number elements also lowers the Auger electron energy.  
% QQQ ref for Auger probability

\begin{figure}[tb]
\centerline{\includegraphics[width=3in]{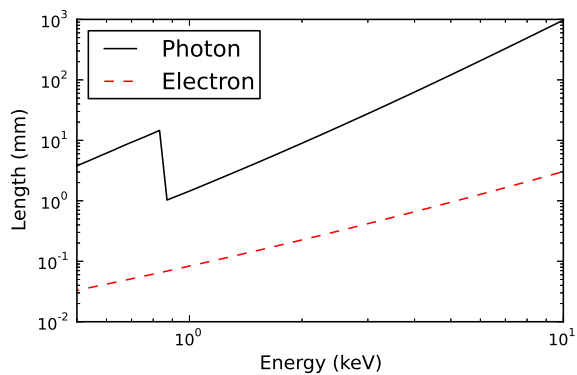}}
\caption{Electron range (dashed line) and X-ray absorption length (solid line) in neon at 1 atm and 0 C.}
\label{neon_range}
\end{figure}

Photoelectric polarimetry can be performed in any detection medium.  However, good modulation factors have been achieved for photoelectric polarimeters only using gas detectors.  The reason is the electron track length.  In silicon, the range of a 1~keV electron is 0.03~$\mu m$, while that of a 10~keV electron is 1~$\mu m$. \cite{Ashley76} Resolving the photoelectron track requires pixels that are a small fraction of electron track length, while solid state X-ray detectors to date have minimum pixel sizes on the order of 10~$\mu$m.  The modulation factors reported for solid state photoelectric polarimeters are all below 10\%.\cite{Bellazzini10}  Increasing the modulation factor would require a decrease in pixel size.  In contrast, the electron range in neon at 1~atm and 0~C is 0.08~mm at 1~keV and 3.0~mm at 10~keV, see Figure~\ref{neon_range}.\cite{Iskef83}  While position resolution on the order of 100~$\mu m$ is feasible in gas detectors, it is quite challenging.

\begin{figure}[tb]
\centerline{\includegraphics[width=4in]{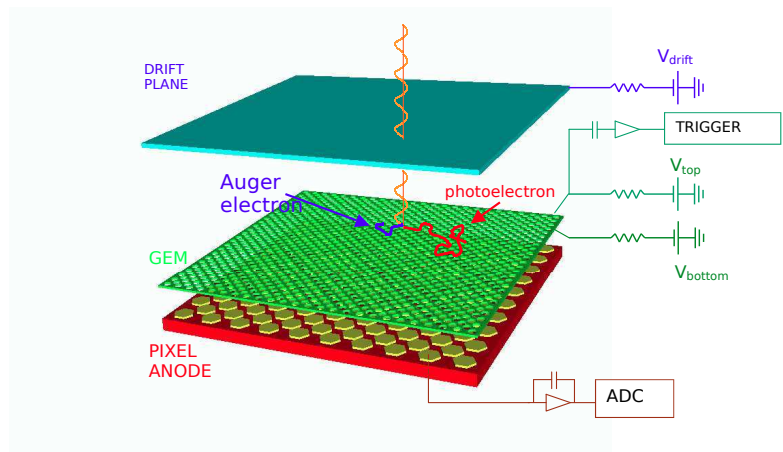}}
\caption{Photo-electric polarimeter with the `Costa' or `parallel-drift' geometry, from Ref.~\refcite{Costa01}.}
\label{pol_costa}
\end{figure}

There are two keys issues in photoelectric X-ray polarimetry with gas detectors: the ratio of photon absorption length to electron track length and the diffusion of the charge carriers in the gas.  Figure~\ref{pol_costa} shows a conceptual view of a gas-filled photoelectric X-ray polarimeter.  X-rays enter at the top of the figure.  To be detected, an X-ray must interact at some point within the gas volume and produce a photoelectron.  The gas volume must be sufficiently deep so that a significant fraction of the X-rays undergo photoelectric interactions.  If the gas layer is too thin, then the detector will have poor quantum efficiency.  The required depth is set by the X-ray attenuation length -- the distance at which $1/e$ of the original X-rays remain.  In neon at STP, the attenuation length is 1.4~mm at 1~keV and 972~mm at 10~keV.  These lengths are much longer than the corresponding electron track lengths, see Figure~\ref{neon_range}.

\begin{figure}[tb]
\centerline{\includegraphics[width=5in]{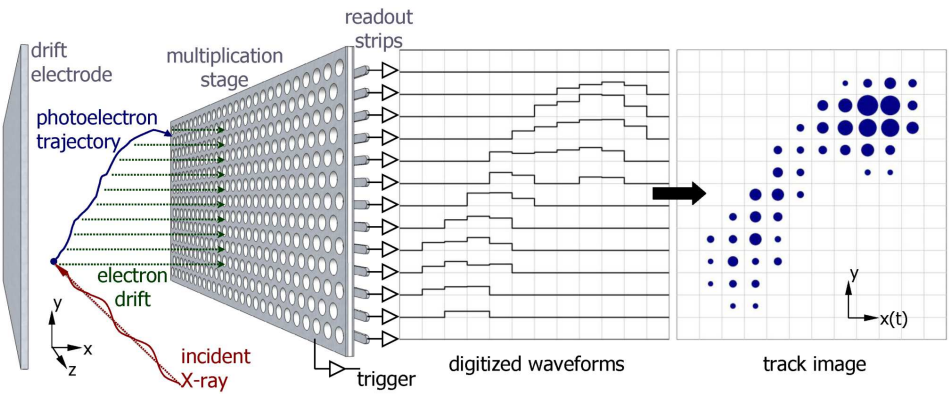}}
\caption{Photo-electric polarimeter with the `Black' or `perpendicular-draft' geometry, from Ref.~\refcite{Black07}.}
\label{pol_black}
\end{figure}

The primary photoelectron produces a track of electron-ion pairs.  The electrons in the track must be brought to readout electrodes at the edge of the detector.  The electrons can be drifted through the gas by application of a uniform electric field.  The drift field can be applied either parallel to the direction of the incident photon, the `Costa geometry' (Fig.~\ref{pol_costa}), or perpendicular, the `Black geometry' (Fig.~\ref{pol_black}).  As the secondary electrons drift, they scatter on the gas atoms.  Thus, localized concentrations of secondary electrons diffuse as they drift.  Diffusion degrades the track image, reducing the accuracy with which the initial track direction can be measured, and reducing the modulation factor.

\subsection{Costa geometry photoelectric polarimeters}

In the Costa geometry (Figure~\ref{pol_costa}), the drift field is applied along the direction of the incident photon.\cite{Costa01}  The photoelectron track is drifted onto a gas electron multiplier (GEM) where it is amplified and then imaged with a two-dimensional array of sensors.  The realization of instruments using the Costa geometry was made possible by the development of the Gas Pixel Detector (GPD) by Bellazzini employing custom CMOS readout electronics fabricated in deep sub-micron VLSI technology.\cite{Bellazzini04,Bellazzini10}.  The latest devices have $\sim$100,000 pixels with 50~$\mu$m pitch covering a 15~mm$^2$ area.\cite{Bellazzini06}  The modulation factor for a detector using this readout device with a 1~cm deep absorption region with 1~atm of 20\% He/80\% dimethyl ether (DME, chemical formula CH$_3$OCH$_3$) has been measured to be 21\% at 2.6~keV, rising to 47\% at 5.2~keV.\cite{Muleri08gpd}  We note that these $\mu$ are quoted with no rejection of events.  Removal of events that are close to circularly symmetric increases the modulation factor at a cost in efficiency.  Allowing the efficiency to decrease to 78\% increases the $\mu$ to 28\% and 54\%, respectively.\cite{Muleri08gpd}

A key advantage of Costa geometry detectors is that they are symmetric under rotation (through multiples of 60$^{\circ}$ for hexagonal pixels) around the incident photon direction.  Measurements using unpolarized X-rays show very low residual modulation, $0.18\% \pm 0.14\%$.\cite{Bellazzini10}  It has been suggested that they can produce accurate polarization measurements without use of rotation.  Another advantage of the Costa geometry is that it provides for true two-dimensional imaging, in addition to polarimetry.  Imaging can be used to lower the instrumental X-ray background for point-like sources and to provide spatially-resolved polarimetry for extended sources.

A disadvantage of the Costa geometry is that the maximum electron drift distance is the same as the maximum X-ray absorption depth.  Since both diffusion and quantum efficiency increase with drift/absorption distance, the Costa geometry requires a trade-off between minimizing diffusion, thus increasing modulation factor, and maximizing quantum efficiency.  The product of quantum efficiency multiplied by modulation factor tends to peak in a relatively narrow band for any specific polarimeter design.

Missions based on Costa geometry polarimeters have been proposed several times.  The most recent is `XIPE: the X-ray imaging polarimetry explorer'.\cite{Soffitta13}
% QQQ add more references 

\subsection{Black geometry photoelectric polarimeters}

In the Black geometry, the drift field is applied perpendicular to the incident photon direction.\cite{Black07}  The photoelectron track is drifted onto a gas electron multiplier where it is amplified and then imaged with a one-dimensional array of sensors.  The second dimension of imaging information is obtained from the time development of the signal on each sensor, thus the detector is a `time projection chamber' (TPC).  This necessitates the use of gases with relatively slow electron drift speeds.  DME has the slowest known electron drift speed.

% developed for a particle physics experiment by David R. Nygren. % ref for TPCs
% and Black geometry detectors are also known as TPC polarimeters

Since the electrons drift perpendicular to the incident photons, the absorption depth is decoupled from the electron drift and large absorption depths can be used.  The absorption depth for the detectors built for the Gravity and Extreme Magnetism Small explorer (GEMS) mission is 31.2~cm.\cite{Hill12}  The GEMS detectors were filled with 190~Torr of DME.  The readout strips had a pitch of 121~$\mu$m and 120 active strips were sampled at a rate of 20~MHz.  The electric field in the gas volume was adjusted to produce a pixel size of 121~$\mu$m on the time axis.  The modulation factor in these detectors was measured to be 29\% at 2.7~keV, rising to 43\% at 4.5~keV.\cite{Hill12}

Use of the Black geometry comes with two costs.  First, the Black geometry uses different techniques to image the two dimensions of the photoelectron track, time versus space.  As noted above, systematic measurement errors are a serious concern in polarimetry and the Black geometry has an intrinsic asymmetry between the two dimensions.  This requires either careful design and operation\footnote{Specifically, careful monitoring and control of the electron drift speed.} of the polarimeter to minimize the asymmetry, rotation of the polarimetry to zero out any net asymmetry, or both.  Measurements using unpolarized X-rays on the GEMS polarimeters showed a residual modulation of $0.21\% \pm 0.28\%$.\cite{Hill12}  

Second, while Costa geometry detectors can image the sky in two dimensions, only one-dimensional imaging of the sky is possible in the Black geometry.  The track image along the time coordinate provides only relative positions of electrons in the track because the overall drift time is unknown.\footnote{If additional instrumentation were added to precisely record the X-ray arrival time, via detection of scintillation photons produced in the initial interaction, then two-dimensional imaging of the sky would be possible.  However, no feasible implementation has been demonstrated.}  The imaging quality of the Black geometry is further degraded if a deep absorption volume is used since the X-rays will be in focus only at one depth and out of focus at all other depths.

While the discussion of photoelectric polarimeters to this point has assumed drift of free electrons, in some gases charge transport occurs via negatively charged ions.  Negative ions offer reduced diffuse and drift speeds compared to electrons.\cite{Martoff05,Martoff09}  This allows larger drift regions and slower electronics (when used in the Black or TPC geometry).\cite{Hill07}  An X-ray polarimeter has been operated using low concentrations of nitromethane (CH$_3$NO$_2$) as the electron capture agent with CO$_2$ providing the balance of the gas.\cite{PrieskornIEEE}  The readout used 120~$\mu$m strips sampled at an effective rate of 167~kHz to produce square pixels with a measured drift velocity of 20~m/s.  The modulation factor was measured to be 38\% between 3.5 and 6.4~keV.\cite{PrieskornPhD}

\section{Compton/Thomson Scattering Polarimeters}

\subsection{Scattering}

At energies above a few tens of keV, Compton scattering is the dominant interaction of X-rays with matter.  When the X-ray energy is an appreciable fraction of the rest mass energy of an electron, the electron will recoil during the interaction, taking energy from the photon.  The cross section is
\begin{equation}
\frac{d\sigma}{d\Omega} = \frac{r_e^2}{2} 
  \left(\frac{E'}{E} \right)^2
  \left(\frac{E'}{E} + \frac{E}{E'} - 2 \sin^2\theta \cos^2 \phi \right)
\end{equation}

\noindent where $r_e$ is the classical electron radius, $E$ is the initial photon energy, $E'$ is final photon energy, and we have averaged over the polarization of the final photon.\cite{Lei97}  The photon energies are related to the scattering angle, $\theta$, as
\begin{equation}
E' = E \left[ 1 + (1- \cos\theta)\frac{E}{m_e c^2} \right]^{-1}
\label{compton_energy}
\end{equation}

For scattering angles near 90$^{\circ}$, the azimuthal distribution of the scattered photon is strongly dependent on the X-ray polarization, thus Compton scattering is effective for polarization analysis.  At low X-ray energies, the electron recoil becomes negligible.  In this limit, known as Thomson scattering, modulation reaches 100\% for 90$^{\circ}$ scattering.

\begin{figure}[tb]
\centerline{\includegraphics[width=3in]{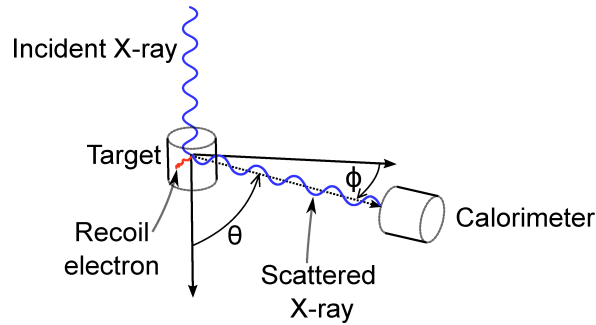}}
\caption{Compton/Thomsom polarimeter.}
\label{compton}
\end{figure}

\subsection{Measurement technique}

The basic principle of all Compton/Thomson polarimeters is shown in Figure~\ref{compton}.  An X-ray scatters on a target.  The scattered X-ray is then detected.  At X-ray low energies, in the Thomson limit, only the scattered photon is detected.  The target/detector geometry is typically arranged to maximize scatterings through polar angles of 90$^{\circ}$ and the detector records the azimuthal distribution of scattered photons.  The target is usually chosen to be a low atomic number material to maximize the ratio of the Thomson versus photoelectron cross section.

If the X-ray is sufficiently energetic, in the Compton regime, it produces a recoil electron.  Thus is it possible to detect both the initial interaction point and the scattered photon.  Compton polarimeters do not require a distinction between target and detector.  Hence, Compton polarimetry is possible in uniform detector arrays.  However, the polarization sensitivity can be improved with the use of low $Z$ targets, since such targets can increase the path length traveled by the scattered photons and also increase the fraction of photons that are Compton scattered rather than photoelectrically absorbed.  In such polarimeters, the target is referred to as an `active target' if recoil electron can be detected and the detector recording the scattered photon is sometimes called a `calorimeter' (since it absorbs the majority of the photon energy).

\subsection{Instruments}

% QQQ check, first?
The first dedicated extra-solar X-ray polarimeter was a Thomson scattering polarimeter flown on a sounding rocket.\cite{Wolff70}  A similar instrument flown later together with a Bragg reflection polarimeter (discussed further below) provided the first successful measurement of the X-ray polarization of an extra-solar object.\cite{Novick72}  

Recently, the Gamma-Ray Burst Polarimeter (GAP) flew aboard the Japanese IKAROS mission.  GAP was designed to measure the polarization of gamma-ray bursts in the 50-300~keV band.  It consists of a single plastic scintillator target (a low $Z$ material) with a diameter of 140~mm surrounded by a cylinder of 12 CsI scintillators.\cite{Yonetoku11}  The modulation factor was measured to be 52\% using an 80~keV pencil beam with 0.8~mm diameter illuminating the center of the target.  Monte Carlo simulations suggest that the modulation factor for astrophysical sources that illuminate the whole target is lower, near 30\% on axis and decreasing off axis.  Uniform response in the CsI scintillators is essential to accurate polarimetry; in-flight calibrations established uniformity at the 2\% level.  GAP detected polarization from three gamma-ray bursts, reporting high average polarizations, 27$\pm$11\% to 84$^{+16}_{-28}$\%, at significances ranging from 2.9$\sigma$ to 3.7$\sigma$ and the detection of variable position angle (at 3.5$\sigma$ confidence) in one GRB.\cite{Yonetoku12}  The systematic uncertainty is dominated by the off-axis response and was estimated to be near 2\% (1-$\sigma$).\cite{Yonetoku11grb}

There are currently several Compton/Thomson polarimeters in various stages of development.\cite{McConnell10}  Several of them use low $Z$ active targets surrounded by high $Z$ calorimeters, specifically the Gamma-RAy Polarimetry Experiment (GRAPE) and the Polarimetry of High ENErgy X-rays (PHENEX) experiment.  PHENEX is a collimated instrument designed to observe known astrophysical sources. GRAPE's primary science goal is GRBs, the but initial balloon flights will use a collimator and point at bright X-ray sources.  

A key issue in these Compton polarimeters is the relatively high background counting rate, which limits the polarization sensitivity.  The light-weight Polarised Gamma-ray Observer (PoGOLite) uses plastic scintillators in the detector and has a large active shield to reduce background.  An even greater reduction in background can be achieved using hard X-ray focusing optics, as demonstrated by the recent success of the NuSTAR mission.  Focusing optics allow use of targets and detectors with greatly reduced volume and a corresponding reduction in background, which can be reduced further via an active target \cite{Chattopadhyay14}.  X-Calibur uses a low $Z$ target surrounded by a CZT detector assembly placed at the focus of a grazing incidence hard X-ray telescope to do polarimetry in the 15-80~keV band.\cite{Beilicke12}.  The detector and shield will rotate around the telescope axis at 10~rpm to minimize systematic effects.  

The wide fields of view needed to catch GRBs preclude use of focusing optics, so hard X-ray GRB polarimeters will necessarily use large detector arrays.  Progress will likely require a dedicated, although potentially small, mission to achieve the total detector volume and mission duration needed to perform polarimetry on a significant sample of GRBs.

\subsection{Measurements with non-polarimeters}

Recently, there have been several polarization measurements using the Compton technique with instruments not designed for polarimetry.  The International Gamma-Ray Astrophysics Laboratory (INTEGRAL) observatory carries the Spectrometer on INTEGRAL (SPI) instrument, which was designed to provide high resolution spectroscopy in the 18 keV to 8 MeV band.  SPI consists of 19 hexagonal Germanium solid-state detectors, surrounded by an anti-coincidence shield, that view the sky through a coded-aperture mask.  To do polarimetry, one selects events in which a gamma-ray deposits energy in two detectors (within a 350~ns coincidence window) and then searches for an azimuthal asymmetry in those detector pairs.  However, other factors, such as the coded aperture shadow pattern and dead detectors within SPI, also affect the pattern of detector pair hits and can produce spurious polarization signatures.  A Monte-Carlo simulation of the instrument can be used to model all of these effects.  Simulations performed with various polarization amplitudes and position angles (varied in addition to the non-polarimetric source parameters such as position on the sky and spectral shape) can then be compared with the observational data obtained on a source and used to estimate the source polarization.\cite{Chauvin13}  Analysis of $5 \times 10^{5}$ double events from the Crab nebula was analyzed via this technique using $7 \times 10^{8}$ simulated events.  The result was a significant detection of polarization in the 0.1-1~MeV band at a level of 46$\pm$10\% at a position angle of 123$^{\circ} \pm 11^{\circ}$.\cite{Dean08}

The measurement has been confirmed using the Imager on Board the INTEGRAL Satellite (IBIS) instrument.  IBIS has two planes of detectors.  Events that trigger one detector in each plane are identified as `Compton events', but only 2\% arise from a true Compton scattering.  IBIS measured a polarization in the 200-800~keV band with a position angle consistent with SPI, but a somewhat higher amplitude.\cite{Forot08}

A number of other measurements have been reported using SPI, IBIS, and the Ramaty High-Energy Solar Spectroscopic Imager (RHESSI), primarily of gamma-ray bursts.\cite{McConnell10}  However, these are of lower significance, for both statistical and instrumental reasons.  Several instruments likely to fly in the next several years, notably the soft gamma detector (SGD) on the Japanese Astro-H mission, will be able to exploit the polarization sensitivity of Compton scattering.  However, instrument not specifically design and operated for polarimetry tend to suffer from instrumental effects that limit their ultimate sensitivity, typically to minimum detectable polarizations (MDPs) on the order of tens of percent.

\begin{figure}[tb]
\centerline{\includegraphics[width=1.25in,angle=-90]{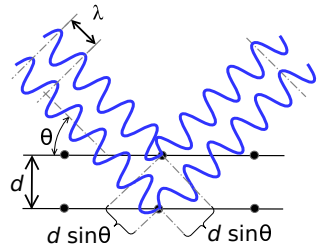}}
\caption{Bragg reflection.  The two outgoing waves are in phase if the difference in path length for scattering from two adjacent crystal planes, $2 d \sin \theta$, where $d$ is the crystal plane spacing and $\theta$ is the angle of incidence, is an integer multiple of the photon wavelength, $\lambda$.}
\label{bragg_refl}
\end{figure}

\section{Bragg Reflection Polarimeters}

At energies below a few tens of keV, X-rays interact more strongly via the photoelectric process than via scattering.  However, superposition of coherent  scatterings off a periodic medium, such as an atomic crystal or multilayer, can produce efficient reflection.  This process is known as Bragg reflection\footnote{The terms Bragg scattering and Bragg diffraction are also used.} and occurs when the difference in path length for scattering from two adjacent crystal planes, $2 d \sin \theta$, where $d$ is the crystal plane spacing and $\theta$ is the angle between the incident ray and the scattering planes is an integer multiple, $n$, of the photon wavelength, $\lambda$, see Figure~\ref{bragg_refl}.  This condition is known as Bragg's law, $n \lambda = 2 d \sin \theta$ or $nhc/E = 2 d \sin \theta$ where $E$ is the photon energy.  

\begin{figure}[tb]
\centerline{\includegraphics[width=3in]{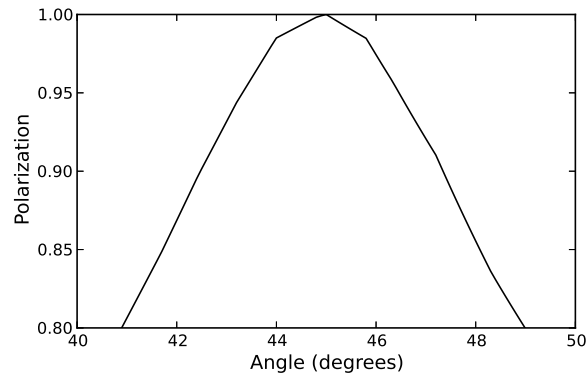}}
\caption{Polarization of 2.6 keV X-rays Bragg reflected off a graphite crystal as a function of incidence angle.}
\label{xtalpol}
\end{figure}

Bragg reflection can be used for polarization analysis because the reflectivity for radiation polarized parallel to the incidence plane is close to zero for incidence angles close to the Brewster angle, which is (very) near 45$^{\circ}$ for X-rays.  The degree of polarization versus incidence angle for 2.6~keV X-rays reflected off a graphite crystal is shown in Figure~\ref{xtalpol}.\cite{Muleri08beams,Henke93}  The modulation factors for Bragg polarimeters are typically very high and can exceed 99\%. Bragg reflection polarimeters must either rotate, to produce a modulation curve as shown in Figures~\ref{analyzer} and \ref{modcurve}, or at least 3 crystals must be used with different position angles (preferably at increments of 45$^{\circ}$) to instantaneously measure the Stokes parameters.
% Brewster angle $\theta_B = \tan$^{-1}$(n_2/n_1)$

Efficient reflection can be obtained for X-rays exactly satisfying the Bragg condition, but the efficiency drops off rapidly as the photon wavelength or incidence angle changes.  The `integrated reflectivity' is the integral of the reflectivity, at fixed energy, over all angles, $\Delta \Theta = \int R(E, \theta) d\theta$.\cite{Silver10}  The effective width is the integral of reflectivity over all energies at fixed angle, $\Delta E(\theta) = \int R(E, \theta) dE$.  The two are related as $\Delta E(\theta_B) = E_B \cot(\theta_B) \Delta \Theta$ where $\theta_B$ is the Bragg angle, usually 45$^{\circ}$ for X-ray polarimeters, and $E_B$ is the corresponding Bragg energy.  The effective width indicates the efficiency of a Bragg polarimeter for an astrophysical source with a broad spectrum.

\subsection{Bragg polarimeters with atomic crystals}

The effective widths of perfect atomic crystals are typically a small fraction of an eV.  Many different crystals\cite{Underwood09} are used for Bragg reflection in laboratory and synchrotron beam experiments, but their effective widths are too small for astronomical applications.  The best effective widths come from ideally imperfect crystals that are a mosaic of small crystal domains with random orientations.  The crystal domains are thin compared with the X-ray absorption length, so an X-ray may pass through multiple domains until it finds one oriented to satisfy the Bragg condition.

The X-ray polarimeter on the OSO-8 satellite used graphite crystals with a mosaic spread of 0.8$^{\circ}$ and an effective width of 3~eV.\cite{Silver10}  Mosaic graphite provides the best effective width in the standard X-ray band (2--10~keV) of any natural crystal.  The OSO-8 polarimeter used a parabolic reflector geometry to focus X-rays onto a small detector in order to minimize the background counting rate.\cite{Novick75}  The range of Bragg angles and the azimuthal extent of each reflector reduced the modulation factor of 0.93.  The OSO-8 instrument contained two orthogonal polarimeters and rotated at a rate of 6~rpm.  Its builders obtained the most precise measurement of X-ray polarization of an astrophysical source to date, showing that the polarization of the Crab nebula at 2.6~keV is 19.2\%$\pm$1.0\% at a position angle of 156.4$^{\circ}$$\pm$1.3$^{\circ}$.\cite{Weisskopf78}

The Stellar X-Ray Polarimeter (SXRP), built for the Soviet Spectrum Roentgen-Gamma mission but never flown, included a Bragg reflection polarimeter using a mosaic graphite crystal in the beam of an X-ray telescope.  The modulation factor was measured to be 99.75\%$\pm$0.11\%.\cite{Tomsick97}  The Astrophysical Polarimetric Explorer (APEX) has been proposed to use parabolic graphite crystal arrays providing a factor of 30 increase in collecting area relative to the OSO-8 polarimeter.  The design has the advantage of a high modulation factor (92.5\%) and the resulting (relative) insensitivity to instrumental effects, but provides measurements only in two narrow bands around 2.6 and 5.2~keV.\cite{Silver10}
% add energy band for OSO-8 due to varying Bragg angle

\subsection{Bragg polarimeters with multilayers}

It is possible to deposit layers of atoms or molecules with thicknesses on the order of nanometers using sputtering or evaporation.\cite{Underwood09}  By depositing alternating layers of high and low atomic number materials, a single high/low $Z$ pair is a `bi-layer', one can manufacture a multilayered structure, or `multilayer', that Bragg reflects.  The Bragg energy is set by the bi-layer thickness and multilayer reflectors are usually best suited for the soft X-ray (below 1~keV) and extreme ultraviolet (EUV) bands.  The reflection efficiency is set by the choice of materials, the number of bi-layers (typically tens to hundreds of layers are needed), and the roughness of both the deposition substrate and of the interface between adjacent layers.  Peak reflectivities above 70\% near normal incidence have been measured for energies near 100~eV, dropping to $\sim$10\% near 500~eV.\cite{Underwood09}  An extensive data base of measured x-ray reflectances for various multilayers is maintained by Lawrence Berkeley National Laboratory the Center for X-ray Optics.\footnote{http://henke.lbl.gov/multilayer/survey.html}  The reflectance of multilayers can also be accurately calculated.\cite{Windt98} 

Multilayer Bragg polarimeters use the same geometries discussed above for crystal polarimeters.  The Polarimeter for Low Energy X-ray Astrophysical Sources (PLEXAS) concept used a parabolic geometry similar to that of the OSO-8 polarimeter, but with a Bragg energy near 250~eV.\cite{Marshall03}. The Bragg Reflection Polarimeter (BRP), that was designed as part of the GEMS mission, used a flat multilayer optic in the beam of one of the GEMS telescopes to provide polarization sensitivity in a narrow band around 500~eV.\cite{Allured13} 

Multilayers offer more flexibility than atomic crystals.  In particular, `graded' multilayers have a varying bi-layer thickness so that the Bragg energy varies across the multilayer surface.  Use of a graded multilayer in a parabolic reflector can compensate for the varying angle of incidence to produce a narrow energy response.  This offers improved background rejection since events outside the energy band can be rejected.

A broad-band soft X-ray polarimeter can be constructed by combining an energy-dispersive grating with a graded multilayer polarization analyzer.\cite{Marshall10}  Gratings, as described in part 3 of this volume, diffract X-rays of different energies through different angles.  A Bragg reflector is highly efficient only at the Bragg energy corresponding to the layer spacing.  By using a graded multilayer, the Bragg energy can be tuned to vary with position to exactly match the energy versus position dispersion of a grating achieving high efficiency across a broad energy range.  If the Bragg reflector is placed at an angle close to 45$^{\circ}$, then it will be a sensitive polarization analyzer.  To obtain a polarization measurement, either the full instrument must rotate to produce a modulation curve or at least 3 different Bragg reflectors must be used with different position angles.  Calculations based on realistic geometries and measured multilayer reflectivities show that modulation factors above 50\% and significant effective area can be achieved across a relatively broad energy band, 200-800~eV.\cite{Marshall14}

\section{Outlook}

Development of new detector and optics technologies has enabled construction of a new generation of astrophysical X-ray polarimeters.  The most exciting advance is the development of high-spatial-resolution gas-filled X-ray detectors and their demonstration as polarimeters exploiting the photoelectric effect.  This technology offers a tremendous increase in efficiency relative to previous devices and should enable polarimetry of a broad range of astrophysical sources.  Broad-band soft X-ray polarimeters based on Bragg reflection are now possible due to advances in the fabrication of multilayer optics via deposition of nanometer thick layers of atoms.  Developments in scintillator hard X-ray detectors has enabled construction of modular, large area Compton scattering instruments suitable for the polarimetry of transient sources requiring large fields of view, while development of pixelated solid-state detectors allows construction of compact hard X-ray polarimeters suitable for use with focusing X-ray telescopes.

\section*{Acknowledgments}

Preparation of this review was greatly aided by the excellent proceedings of the meeting \emph{X-ray Polarimetry: A new Window in Astrophysics} held in Rome in 2009.  Readers seeking further information should first consult this volume.  I thank Martin Weisskopf and Hannah Marlowe for their comments that improved the manuscript and Kevin Black and Enrico Costa for useful discussion and providing figures.  I acknowledge intermittent funding support from NASA for X-ray polarimetry.

%\section{Bibliography}\label{secbib}\index{bibliography}

%\printindex[aindx]                 % to print author index
%\printindex                         % to print subject index
\end{document}